\input harvmac\skip0=\baselineskip

\lref\gvw{B.~Greene, C.~Vafa, N.~Warner,``Calabi-Yau Manifolds and
Renormalization Group Flows,'' Phys. Lett. {\bf 218B}, 51 (1989).}
\lref\mms{
  M.~Marino, R.~Minasian, G.~W.~Moore and A.~Strominger,
 ``Nonlinear instantons from supersymmetric p-branes,''
  JHEP {\bf 0001}, 005 (2000)
  [arXiv:hep-th/9911206].}
\lref\msw{J.~Maldacena, A.~Strominger, E.~Witten, ``Black Hole Entropy in
M-Theory,'' JHEP {\bf 9712}, 002 (1997) [arXiv:hep-th/9711053]}
\lref\curtjuan{C.~Callan, J.~Maldacena, ``Brane Dynamics from the
Born-Infeld Action,'' Nucl. Phys. {\bf B513} 198-212 (1998)
[arXiv:hep-th/9708147]}
\lref\fw{D.~Freed, E.~Witten, ``Anomalies in String Theory with
D-Branes,'' [arXiv:hep-th/9907189]}
\lref\gsy{D.~Gaiotto, A.~Strominger, X.~Yin,``Superconformal Black Hole
Quantum Mechanics,'' [arXiv:hep-th/0412322]}
\lref\VafaGR{
  C.~Vafa,
  ``Black holes and Calabi-Yau threefolds,''
  Adv.\ Theor.\ Math.\ Phys.\  {\bf 2}, 207 (1998)
  [arXiv:hep-th/9711067].
}
\lref\cvw{
  B.~R.~Greene, C.~Vafa and N.~P.~Warner,
  ``Calabi-Yau Manifolds And Renormalization Group Flows,''
  Nucl.\ Phys.\ B {\bf 324}, 371 (1989).
}

\lref\Martinec{
  E.~J.~Martinec,
  ``Algebraic Geometry And Effective Lagrangians,''
  Phys.\ Lett.\ B {\bf 217}, 431 (1989).
}
\lref\ShihQF{
  D.~Shih, A.~Strominger and X.~Yin,
  ``Counting dyons in N = 8 string theory,''
  arXiv:hep-th/0506151.
}

\lref\GreeneDenef{
  F.~Denef, B.~R.~Greene and M.~Raugas,
  ``Split attractor flows and the spectrum of BPS D-branes on the quintic,''
  JHEP {\bf 0105}, 012 (2001)
  [arXiv:hep-th/0101135].
}

\lref\nakajima{
H.~Nakajima, {\sl Lectures on Hilbert Schemes of Points on Surfaces},
American Mathematical Society, Providence, RI, 1999. }

\lref\DG{M.~Douglas and G.~Moore, ``D-branes, Quivers, and ALE
  Instantons,''
[arXiv:hep-th/9603767]}
\lref\Denf{F.~Denef, ``Quantum Quivers and Hall//Hole Halos,''
JHEP{\bf 10}, 023 (2002)[arXiv:hep-th/0206072]}
\lref\Gep{A.~Recknagel and V.~Schomerus, ``D-branes in Gepner models,''
Nucl. Phys. {\bf B531}, 185-225 (1998) [arXiv:hep-th/9712186]}
\lref\Kth{R.~Minasian and G.~Moore, ``K-theory and Ramond-Ramond
Charge,'' JHEP {\bf 11}, 002 (1997) [arXiv:hep-th/9710230] }
\lref\Wit{E.~Witten, ``D-branes and K-theory,'' JHEP {\bf 9812}, 019
(1998) [arXiv:hep-th/9810188]}
\lref\Har{M.~B.~Green, J.~A.~Harvey and G.~W.~Moore, ``I-brane inflow
and anomalous couplings on D-branes,'' Class Quant. Grav. {\bf 14}, 47
(1997) [arXiv:hep-th/9605033]}
\lref\Che{Y.~K.~Cheung and Z.~Yin, ``Anomalies, branes, and currents,
'' Nucl. Phys. {\bf B517}, 69 (1998) [arXiv:9710206] }
\lref\WiPh{E.~Witten, ``Phases of $N=2$ Theories in Two Dimensions,''
Nucl. Phys. {\bf B403}, 159-222 (1993) }
\lref\Bru{I.~Brunner, M.~R.~Douglas, A.~Lawrence, C.~Romelsberger,
``D-branes on the Quintic,'' JHEP {\bf 0008}, 015 (2000)[arXiv:hep-th/9906200]}
\lref\aspone{
  P.~S.~Aspinwall,
  ``D-branes on Calabi-Yau manifolds,''
  arXiv:hep-th/0403166.
}
\lref\asptwo{
  P.~S.~Aspinwall and S.~Katz,
  ``Computation of superpotentials for D-Branes,''
  arXiv:hep-th/0412209.
}

\lref\bott{R.~Bott and L.~Tu, {\sl Differential Forms in Algebraic
Topology},
Graduate Texts in Mathematics, {\bf vol.82}, Springer-Verlag, New York, 1982. }

\lref\dsw{M.~Dine, N.~Seiberg, E.~Witten, Nucl. Phys. {\bf B289}, 589,
(1987)}
\lref\sw{N.~Seiberg, E.~Witten, JHEP {\bf 9909}, 032 (1999)}

\lref\aspthree{
  P.~S.~Aspinwall and L.~M.~Fidkowski,
  ``Superpotentials for quiver gauge theories,''
  arXiv:hep-th/0506041.
}
\lref\mrdthree{
  M.~R.~Douglas, S.~Govindarajan, T.~Jayaraman and A.~Tomasiello,
  ``D-branes on Calabi-Yau manifolds and superpotentials,''
  Commun.\ Math.\ Phys.\  {\bf 248}, 85 (2004)
  [arXiv:hep-th/0203173].
}
\lref\mrdtwo{
  M.~R.~Douglas,
  ``Lectures on D-branes on Calabi-Yau manifolds,''
{\it Prepared for ICTP Spring School on Superstrings and Related Matters, Trieste, Italy, 2-10 Apr 2001}
}\lref\mrdone{
  I.~Brunner, M.~R.~Douglas, A.~E.~Lawrence and C.~Romelsberger,
  ``D-branes on the quintic,''
  JHEP {\bf 0008}, 015 (2000)
  [arXiv:hep-th/9906200].
}

\Title{}{D4-D0 BRANES ON THE QUINTIC}

\centerline{ Davide Gaiotto, ~Monica Guica, ~Lisa Huang, }
\centerline{ Aaron Simons, ~Andrew Strominger and Xi Yin}
\smallskip
\centerline{Jefferson Physical Laboratory, Harvard University,
Cambridge, MA 02138} \vskip .6in \centerline{\bf Abstract} { It is proposed that the quantum mechanics of
$N$ D4-branes and $M$ D0-branes on the quintic is described by the
dimensional reduction of a certain
$U(N)\times U(M)$ quiver gauge theory, whose superpotential encodes the defining quintic polynomial.
It is shown that the moduli space on the Higgs branch exactly reproduces the moduli space of degree $N$
hypersurfaces in the quintic endowed with the appropriate line bundle, and that the cohomology growth
reproduces the D4-D0 black hole entropy.  }
\vskip .3in

\smallskip
\Date{September 21, 2005}

\listtoc \writetoc
\newsec{Introduction}

BPS black holes in IIA compactification on a  Calabi-Yau threefold
$X$  are constructed by wrapping D-branes around even-dimensional
cycles.  In general one expects the brane quantum mechanics, which
should be the large N dual of the black hole, to be a kind of
dimensional reduction of an $N=1, ~~d=4$ quiver gauge theory, with
one node for each type of cycle \refs{\DG \mrdone \GreeneDenef \Denf \mrdtwo
\mrdthree \aspone \asptwo - \aspthree}.   On the other hand, the
microstates of many such black holes have been identified with
moduli space cohomology of high degree subvarieties in $X$ \refs{\msw,\VafaGR}.
Hence one expects a direct connection between the moduli spaces of
quiver gauge theories and of  subvarieties in $X$. It is the
purpose of this paper to study this problem and, more generally,
the dual relation between BPS black holes and quiver gauge
theories.

We focus herein on the canonical example of the $N$ D4-branes
wrapping the basic  hypersurface in the quintic, denoted $P_1$, and
$M$ D0-branes  on the quintic.\foot{Extensive analyses of exact D-brane boundary
states, not considered here, can be found in \refs{\mrdone,\Gep}. }  $P_1$ turns out not to be a spin
manifold and hence a
wrapped D4-brane is required to have a gauge field valued in
half-integral cohomology  and an induced D2 charge \refs{\Kth \Che  \fw -\Wit}.
 This system is
relatively simple but still rich enough to describe macroscopic
black holes. We analyze the
field content, superpotential and D-term constraints of the
low-energy D-brane gauge theory, which should be enough to capture
the topological features of the theory. We propose that it is (the
reduction from $d=4$ to $d=1$ of) a $U(N)\times U(M)$ quiver gauge
theory with 4 chiral multiplets in the adjoint of $U(N)$,
corresponding to the 4 deformations of $P_1$, 3 chiral multiplets
in the adjoint of $U(M)$, corresponding to the 3 motions of a D0
in the quintic, and 2 bifundamentals, together with a certain superpotential
connecting all 3 types of chiral multiplets.

 For $M=0$, (no D0-branes) one expects that  the moduli space of the Higgs branch should
match  the moduli space of degree $N$ hypersurfaces, denoted
$P_N$.  At first sight this seems impossible, because the dimension of the latter grows
like $N^3$, while the gauge theory contains only order $N^2$ fields.
However it is important to realize that the $P_N$  obtained by combining
$N$ $P_1$'s
must come  with  a particular supersymmetric  $U(1)$ worldvolume gauge field $F$ as required  by D0
and D2 charge conservation. We show that the existence of a suitable such $F$ is equivalent to the
existence of a suitable holomorphic divisor on $P_N$.  Requiring the existence of such a divisor reduces the
dimension of the moduli space from a number
which grows like $N^3$ to one that grows like $N^2$.    Indeed we find a
remarkable match between the moduli space of such $P_N$'s and the moduli
space of the gauge theory Higgs branch. We further study the problem of
combining hypersurfaces of different degrees and gauge fields and  find perfect agreement with the
appropriate gauge theory Higgs branch.

We then show
that the coupling of $M$ D0-branes to the associated $U(M)$ gauge
sector leads to a factor in the Higgs branch moduli space
involving the symmetric product of $M$ copies of $P_N$. This
factor has a large cohomology, leading to a quantum ground state
degeneracy which reproduces the Bekenstein-Hawking area-entropy
law $S=2\pi\sqrt{5N^3M}$.

Some interesting features emerge from our analysis which we expect
to carry over to a generic Calabi-Yau. The quiver gauge quantum
mechanics is expected to be dual, at least at the level of
topological data, to string theory on the corresponding attractor geometry. The
K{\"a}hler moduli of the attractor geometry are fixed in terms of the
black hole charges- in this case $(N,M)$-while the complex moduli
can be continuously varied. This is mirrored in the gauge theory,
where the superpotential depends only on the complex structure and
can be continuously varied, while the charges $(N,M)$ determine the
gauge symmetry and field content.

A natural question which arises is the relation of the quiver
quantum mechanics discussed here and the superconformal quantum
mechanics expected from $AdS_2/CFT_1$ -- for example the one
discussed in \gsy.  We don't know the answer to this interesting
question, though it may involve some kind of RG flow and
integrating out of fields.

This paper is organized as follows. In section 2 we study the
classical geometry of degree $N$ hypersurfaces with gauge fields
and curvatures inducing D0 and D2 charges, and how several such
cycles may be joined in a manner dictated by charge conservation.
In section 3 we show that this geometric structure is exactly
reproduced by the Higgs branches of appropriate quiver gauge
theories. In section 4 we match the ground state degeneracy of
the quiver gauge theory to the black hole entropy. In section 5
we study the geometry of joining a general set of D4-branes
in a manner that is consistent with charge conservation.

In the text we will consider both space-filling wrapped branes
and those which are pointlike in the noncompact space. While
ultimately we are interested in the latter the
two are related by dimensional reduction and the former are
often simpler and more familiar. In either case we will refer
to the brane by the real dimension of the cycle which it wraps.
We will also use symbols such as $F$, $J$,
$P$ or $\Sigma_A$ to denote either a cycle or its dual form, with
the precise meaning hopefully clear from the context.

\newsec{Geometry}

\subsec{Hypersurfaces in $X$}
In this section we review some basic facts about single D4 branes wrapping
a degree $N$ hypersurface $P_N$ in the quintic $X$.

The Chern class of  $X$ is
\eqn\bgk{c(TX)=1+10J^2-40J^3}
with $J$ the hyperplane section and $\int_XJ^3=5$. The Chern class of a degree $N $
divisor $P=NJ$ is
\eqn\cfh{c(TP)=1-NJ+(10+N^2)J^2,}
with $\int_P J^2 =5N$.
$P_N$ has Betti numbers
\eqn\dza{b_0=1,~~b_1=0,~~b_2^+={5N^3+25N \over  3}-1,~~~b_2^-={10N^3+125N \over 3}-1}
so that the Euler character is $\chi =5N^3+50N$ and the signature is $\sigma=-(5N^3+100N)/3$.
The complex moduli space ${\cal M}(P)$ consists of the $\half(b_2^+-1)$ complex deformations of $P_N$ in $X$.

$P_N$ may also carry a  two-form $U(1)$ field strength $F$. This and the curvature of $TP_N$
induce a D0 charge \Kth\
\eqn\rtop{q_0=-\int_{P_N} \bigl( \half F^2+{1 \over 24}c(TP_N)\bigr)=k -{5N^3+50N \over 24},}
where $k$ is the instanton number. There is also induced  $D2$ charge
\eqn\indd{q_2=\int_{P_N} F \wedge J.  }

Supersymmetry requires that $F$ take the form\foot{For large $F$ these equations
must be deformed \mms . We expect this will not qualitatively affect our discussion.}
\eqn\dxi{F=F^-+{q_2 \over 5N}J,}
with
\eqn\rti{F^-=-*F^- .}
This is equivalent to saying $F^-$ is orthogonal to $J$ and is of type $(1,1)$.
Furthermore, for $N$ even $F$ must be in integer cohomology, but for
$N$ odd it must be shifted by $\half J$, so that $q_2$ and $q_0$ are always nonzero and fractional \refs{\Kth-\Wit}.
Note that \rti\ depends on the complex structure of $P_N$ (as induced from $X$). Hence in general a solution of \rti\ does not remain a solution under deformations of $P_N$ in $X$ and moduli are frozen
when $F^-$ is nonzero.

Suppose two surfaces of degrees $N'$ and $N''$and instanton numbers $k'$ and $k''$ join into a single one of degree
$N'+N''=N$.
$D0$ charge conservation requires the final instanton number
\eqn\raz{k =k'+k'' +{5N'^2 N''+5 N' N''^2 \over 8} . }

\subsec{$N=2$}
Now we try to combine two degree one $P_1$ surfaces to get a single
degree 2 surface $P_2$.  The initial surfaces have $F=\half J$ so that
the total charges are  \eqn\tik{2q_2(1)=5,~~~~2q_0(1)=-{35 \over
6}.} This implies that the degree 2 surface must have
\eqn\raz{k=-\half \int_{P_2} F^2=0,}\eqn\fdx{\int_{P_2} F\wedge J=5,} with
$F=F^-+\half J$ an integral form. Since $F$ is also type $(1,1)$
it can be represented by a holomorphic divisor, in which
case \raz\ and \fdx\ represent intersection numbers (we also use the symbol
$F$ to denote this divisor). At generic points there are no
divisors other than $J$, so we cannot make a generic degree 2
surface by combining two degree 1 surfaces. However consider surfaces described by
quadratic equations of the form \eqn\rfv{{\rm det}\Phi=0,} where $\Phi=\Phi_Az^A$ is a
$2\times 2$ matrix of linear polynomials in $z^A$, $A=1,\dots, 5$.
There is a 13-dimensional moduli space of such surfaces. They
admit a divisor $F$ described by the non-transverse equations
\eqn\fto{\Phi_{11}=0,~~~\Phi_{12}=0.} To compute \fdx\  we note
that $J$ and \fto\ intersect at  5 points in the quintic. Since
these 5 points automatically lie in the surface \rfv\ this is also
the intersection of $F$ with $J$ . To compute the self
intersection of $F$, we note that \fto\ may be deformed
to\eqn\ftx{\Phi_{11}=\epsilon \Phi_{21},~~~\Phi_{12}=\epsilon
\Phi_{22}.} This clearly has no intersection with the original
curve, in agreement with \raz.

This agrees with the number of Higgs branch moduli of a $U(2)$ gauge theory with four
chiral multiplets in the adjoint, one for each deformation of $P_1$ in $X$. On the Higgs branch with
$U(2)\to U(1)$, three chiral multiplets are eaten by the Higgs
mechanism. The remaining 16-3=13 neutral chiral multiplets will be found in section 3 to
match
the 13 moduli of the degree 2 surface with $F$ described above.

\subsec{$N=3$}

Next we combine  three degree 1 surfaces into a degree 3 surface $P_3$.
 The initial surfaces  again have $F=\half J$ and
the total charges are  \eqn\tikv{3q_2(1)={15 \over 2},~~~~3q_0(1)=-{35 \over
4}.} This implies that the degree 3 surface must have
\eqn\razv{k=-\half \int_{P_3} F^2={25 \over 8},}\eqn\fdxv{\int_{P_3} F\wedge J={15 \over 2},} with
$F=F^-+\half J$.  For $N$ odd, $F^-$ (rather than $F$) is a type $(1,1)$
integral
form and can be represented by a divisor of $P_3$ on a subvariety of the moduli space.
This subvariety is described by cubic
equations of the form \eqn\rfv{{\rm det}\Phi=0,} where $\Phi$ is now a
$3\times 3$ matrix of linear polynomials in $z^A$.
There is a 28-dimensional moduli space of such surfaces. They
admit a divisor $C$ described by the non-transverse equations
\eqn\fpo{C:~~~~~~~~~~~\tilde \Phi_{11}=0,~~~\tilde \Phi_{12}=0,~~~\tilde \Phi_{13}=0,} where
\eqn\trov{  ~~~\tilde \Phi_{kk'}=\half \epsilon_k^{~~ij}\epsilon_{k'}^{~~i'j'}\Phi_{ii'}\Phi_{jj'}}
is the minor associated to $\Phi_{kk'}$.  $C$ intersects $J$ at 15 points, as we show below in section 2.4.   $C$ may be continously deformed to
\eqn\deza{v^j{\tilde \Phi_{jk}=0}}
for any nondegenerate vector $v^k$. To compute the self intersection of  $C$ take the deformed surface defined by  $v=(0,0,1)$.
This intersects \fpo\ at
\eqn\ftov{~~~~~~~~~~~ \Phi_{21}=0,~~~\Phi_{22}=0,~~~\Phi_{23}=0,}
which consists of  5 points in the quintic.

$C$ is not anti-self-dual since it intersects $J$.  An anti-self-dual form is defined by
\eqn\fxa{F^-=C-J .}
so that
\eqn\wsa{F=C-\half J.}
It is readily checked that $F$ obeys \razv\ and  \fdxv\ as required.

This concurs with our expectation from a $U(3)$ gauge theory with four
chiral multiplets in the adjoint. On the Higgs branch with
$U(3)\to U(1)$,   8 chiral multiplets are eaten by the Higgs
mechanism. The remaining 36-8=28 neutral chiral multiplets match
the 28 moduli of the degree 3 surface with $F$ described above.

\subsec{General $N$}

Consider the complex hypersurface in the quintic defined by
\eqn\detqu{ P_N:~~~~\det \Phi=0 }
where $\Phi$ is an $N\times N$ matrix of linear combinations of the $z^A$'s.
Since a  complexified $SU(N)$ transformation of $\Phi$ leaves the hypersurface unchanged,  such hypersurfaces
have a moduli space of complex dimension
$3N^2+1$.  This contrasts with general degree $N$ hypersurfaces, which have a dimension
${5N^3+25 N \over 6}-1$. We note that the moduli space of hypersurfaces \detqu, unlike the more general case,
does not depend on the
defining polynomial for the quintic.  These hypersurfaces are also special in that they  admit a holomorphic
curve $C$ defined by the vanishing of the first (for example) row of minors
of $\Phi$
\eqn\csd{ C: ~~~~ \tilde\Phi_{1i}=0,~~~1\leq i\leq N. }
We could consider curves defined by other linear combinations of rows
of minors, but they are homologous to $C$ since they can be continuously
deformed to $C$.\foot{We could also consider a column of minors, which gives a surface
homologous to $C$ plus a multiple of $J$.} Note that \csd\ is a set of non-transverse equations.

Let us compute the intersection number $C\cdot J$. We can represent $J$
by the hyperplane $H(z)=\sum h_A z^A=0$. Setting the first row of
minors of $\Phi$ to zero is equivalent to the $N-1$ rows $(\Phi_{ik})_{1\leq k\leq N}$
being linearly dependent for $i=2,\dots,N$. Now it amounts to counting
the number of solutions to the set of equations
\eqn\seqs{ \eqalign{ & U(z^A)=0, ~~~~H(z)=0, \cr &\sum_{2\leq i\leq N}
c_i \Phi_{ij}(z) =0,~~~~1\leq j\leq N, } } on ${\bf P}^4\times {\bf
P}^{N-2}$, where
$c_i$'s are the homogeneous coordinates on ${\bf P}^{N-2}$, and $U$ is the defining degree 5 polynomial for the quintic.
Since the curve given in \csd\ is not a complete intersection, we have introduced a
set of auxiliary variables $c_n$ in ${\bf P}^{N-2}$, and a set of
equations which
define the curve as a complete intersection. Now we can use standard techniques to compute
the intersection $C\cdot J$,\foot{ The calculation
amounts to counting the zeros of a section (i.e. the Euler class) of the vector bundle $V=i_1^* H_1\oplus
i_1^* H_1^{\otimes 5}\oplus (i_1^* H_1\otimes i_2^* H_2)^{\oplus N}$,
where $H_1$ and $H_2$ are the hyperplane bundles over ${\bf P}^4$ and ${\bf P}^{N-2}$
respectively, $i_1$ and $i_2$ are the natural projection maps from their product. }
\eqn\cdj{\int_{{\bf P}^4 \times {\bf P}^{N-2}}
(1+x)(1+5x)(1+x+y)^N=5 {N \choose 2}.}
Next we will compute the self-intersection $C \cdot C$. This amounts to setting
the first two rows of minors of $\Phi$ to zero, which is equivalent to
having the $N-2$ rows $(\Phi_{ik})_{1\leq k\leq N}$ being linear dependent,
$i=3,\cdots,N$.
So we can describe the points in $C\cdot C$ by the set of equations
\eqn\eqas{ \eqalign{ &U(z^A)=0, \cr & \sum_{3\leq i\leq N}
c_i\Phi_{ij}(z)=0,~~~ 1\leq j\leq N. } }
The number of solutions is given by
\eqn\cdc{\int_{{\bf P}^4 \times {\bf P}^{N-3}} (1+5x)(1+x+y)^N=
5{N \choose 3}.}

We shall identify the curve $C$ with an integral $(1,1)$ harmonic form on $P_N$.
We can construct an anti-self-dual form
\eqn\ansd{ F^- = C-{N-1\over 2}J }
and the flux
\eqn\fffl{ F = {J\over 2}+F^- }
The total induced D2-brane charge on the D4-brane wrapped on $P_N$ is
\eqn\qdteo{ q_2 = F\cdot J = {5\over 2}N, }
and the total induced D0-brane charge is
\eqn\qdzero{ \eqalign{ q_0 & = -\int_{P_N} {F^2\over 2} +{c_2(TP_N)\over 24} \cr
&= -{35\over 12}N } }
These are precisely $N$ times the charges of a D4-brane wrapped on $P_1$
with $F=J/2$.

\newsec{Gauge Theory}

\subsec{$N$ D4-branes}
In this subsection we consider a stack of $N$ D4-branes all wrapped
on the same hypersuface  $P_1$, given by the linear equation $\phi_Az^A=0$ in the quintic,
with the minimal required flux $F=\half J$ on each. This is
described at low energies in the noncompact spacetime by the
dimensional reduction (to one dimension) of a four-dimensional
${\cal N}=1$ $U(N)$ gauge theory.  In general this theory contains
an infinite number of higher dimension operators and cannot be
described exactly. However here we are interested only in the
topological properties which are encoded in the $F$ and $D$ terms. On general grounds \aspone\  it is expected that the
$F$ terms depend only on the complex structure while the $D$ terms depend
only on the K{\"a}hler class.

In addition to a $U(N)$ gauge multiplet the theory contains 4
adjoint chiral matter fields, corresponding to the 4 complex deformations of the
surface $P_1$ in the quintic. The latter live in the projective space ${\bf P}^4$, with
projective coordinates $\Phi_A$, promoted to $N\times N$ matrices.
These are modded out by a $GL(N,{\bf C})$ action \eqn\glncs{ \Phi_A
\to g \Phi_A,~~~~g\in GL(N,{\bf C}) ,} which reduces the 5
$\Phi_A$ to 4 physical fields. It is easiest to restrict to an
affine patch, on which  $\Phi_1$  can be set to the identity matrix and the
remaining $\Phi_A$ $A=2,\dots ,5$ are independent $N\times N$
matrices. The moduli space of the Higgs branch is determined by the D-term
constraint \eqn\dterms{ G^{A\bar B}[\Phi_A,(\Phi^\dagger)_{\bar
B}]=0, } which consists of $N^2-1$ real equations.  The precise form of the moduli space metric
$G^{A\bar B}$ will not be needed. Modding out by the $SU(N)$ gauge action
(the $U(1)$ acts trivially) leads to a  $3N^2+1$ dimensional complex moduli space.

\subsec{Adding one D0-brane}

Now we add a single D0-brane.
This gives a $U(N)\times U(1)$ quiver gauge theory, with
bifundamental matter $\phi_{04},\phi_{40}$ coming from 0-4 and 4-0
strings, and neutral matter $Z^A$ associated to the coordinates of
the D0-brane in the CY. Again the $Z^A$'s are naturally projective
coordinates, but we will work on the affine patch and set $Z^5=1$.
There is a superpotential \eqn\suppo{ W = \phi_{04}
\Phi_{A} \phi_{40} Z^A + \Lambda
U(Z^A) } which involves only the complex structure.
 Here we have introduced a chiral superfield Lagrange
multiplier $\Lambda$ to impose the condition that the $Z^A$'s lie on the
quintic defined by $U(Z^A)=0$.\foot{It is likely there is a more elegant formulation following the Landau-Ginzburg construction of Calabi-Yau
sigma-models \refs{\gvw,\Martinec,\WiPh}. For now we take a more pedestrian approach.}
The non-zero D2 brane charge $q_2=5N/2$ leads to a $U(1)$
Fayet-Illiopoulos term and associated D-term constraint\foot{This is discussed in a slightly
different language in \msw.  For a surface $p^A\Sigma_A$ with flux $F=f^A\Sigma_A$ on a more general Calabi-Yau with Kahler class
$J^A\Sigma_A$, the right hand side wold be proportional to
$D_{ABC}p^AJ^Af^A$. This involves only even classes, as befits a
$D$-term.} \refs{\DG,\dsw,\sw}
\eqn\ewds{\phi^\dagger_{40}\phi_{40}-
\phi_{04}\phi^\dagger_{04}=5N\zeta.}
The D-term constraint \dterms\ is also deformed.
\ewds\ forces a 4-0 vev and breaks the
$U(1)\times U(1)$ symmetry (associated to the D4 and D0 centers of mass)
down to $U(1)$.  Once these vevs are nonzero, generic values of
the D0-brane coordinate $Z^A$ no longer minimize the potential. This
corresponds to the fact that
the presence of D2 charge on the D4 leads to an attractive force between
the D4's and the D0-brane.

  We must also ensure that the the superpotential is stationarized. This requires that $\phi_{40}$ and $\phi_{04}$ are zero eigenvectors of   $\Phi_Az^A$.  This is possible if and only if   the latter has a zero eigenvalue, i.e.
\eqn\ddx{{\rm det}\bigl[\Phi_AZ^A \bigr] = 0.} \ddx\ is obeyed.
Hence we precisely  recover the hypersurface equation
derived earlier from geometry in \detqu, from the D and F flatness conditions of the quiver
gauge theory!\foot{
For generic values of $Z^A$, ${\rm det}\Phi_AZ^A \neq 0$ and the first term in \suppo\ gives masses to
all the 0-4 and 4-0 strings.  This corresponds to the fact that the if the D0 and D4 are not coincident
these strings are stretched and massive. Hence \ddx\ may also be viewed, along the lines of \curtjuan\
as the locus of massless strings connected to a D0 probe.}

\subsec{$M$ D0-branes}
More generally we can consider having $M$ D0-branes and $N$
D4-branes wrapped on the cycle $J$. Then we have a $U(N)\times
U(M)$ gauge theory with D0-brane coordinates $Z^A$ in the adjoint of
$U(M)$, D4-brane moduli $\Phi_{A}$ in the adjoint of
$U(N)$. The key terms (up to higher order commutators) in the superpotential are \aspone
\eqn\supww{ W = {\rm Tr}\left[ \phi_{04} \Phi_{A}\phi_{40}Z^A\right] + {\rm Tr} \Lambda U(Z^A)
+ {\rm Tr} \,\Omega_{ABC} Z^A[Z^B,Z^C], } with $\Omega$ the holomorphic  three form.\foot{This can be written as the pullback from $P^4$ of $\epsilon_{ABCD}{\partial z^D \over \partial U}$ for $A,B=1,...4$.} We are interested in the branch
on which all the D0-branes coincide with the D4-brane world volume,
corresponding to \eqn\detac{ {\det}_N\left[ \Phi_{A} Z^A \right]=0, } where the
determinant is only taken with respect to the $U(N)$ indices of
$\Phi_{A}$, and with $\phi_{04}=0$, and $\phi_{40}$ a zero eigenvector of
$\Phi$.\foot{There may be other branches --- in fact some might be expected corresponding to D0 branes dissolving into $F^-$.} This is the defining equation for $P_N$ with $F=\half J$, but with $z^A$'s replaced by $M\times M$ matrices $Z^A$.
The F-term
constraints are solved by taking \eqn\supcon{ U(Z^A)=0,~~~~[Z^A,Z^B]=0. } Note that
despite the ordering ambiguities in \supww, the condition \supcon\
consistently solves the supersymmetry constraints since all $Z^A$'s
commute with one another. Together with \detac\ and the D-term constraints,
these define the Hilbert scheme of $M$ points on $P_N$ \nakajima,
corresponding to the moduli space of $M$ pointlike instantons on the
D4-brane world volume.

\newsec{D4-D0 Black Hole Entropy}
Given the preceding construction of the classical moduli space
of the quiver gauge theory, the problem of counting the ground
states to leading order and matching to the black hole entropy is essentially
equivalent to previous analysis \refs{\msw,\VafaGR, \gsy}. Nevertheless we give a brief review for completeness.
The large entropy for large $(M,N)$ arises from the D0's. For fixed $P_N$,
their BPS states are in one-to-one correspondence with
the cohomology classes of the symmetric product orbifold  ${\rm Sym}^{M}(P_N)$.
These can be constructed from the basic cohomology classes of
$P_N$
and their counterparts appearing in the $M$ twisted
sectors of the orbifold. The cohomology classes  can be organized as
\eqn\chiprim{\prod_{i=1}^k \alpha_{-n_i}^{A_i}|0\rangle.} The $ n_i$ take
values from 1 to $M$ (labeling the twisted sector) with the
restriction that the sum equals $M:$
\eqn\cst{\sum_{i=1}^k n_i=M.} The index $A=0,1,\dots ,5N^3+50N$ runs
over the $\chi(P)$ cohomology classes with  $A=0$ corresponding to
$H^0(P)$. The state
\eqn\frt{(\alpha^0_{-1})^{M}|0\rangle} corresponds to  $H^0({\rm
Sym}_{M}(P_N))$.
The counting of such states is equivalent to the counting of left-moving states of a
$c_L=\chi(P_N)$  2D CFT at level $M$. We accordingly find the asymptotic formula for the entropy
\eqn\rfz{S=2\pi\sqrt{5N^3M}.}
This agrees with the Bekenstein-Hawking area law for the corresponding extremal charge $(N,M)$ black hole.

\newsec{D4-branes on Surfaces of Higher Degrees}

In this section we consider the geometry of D4-branes wrapped on
higher degree surfaces, with the proper fluxes turned on. As in
section 2, when the D4-branes join together, charge conservation requires
a certain integral $(1,1)$ flux that exists only on a special
class of hypersurfaces. In the case when all the D4-branes are wrapped on the
same cycle, we find a gauge theory description whose moduli space
matches that of the above mentioned hypersurfaces with flux.
For the more general case of D4-branes wrapped on
different 4-cycles, we propose a geometric construction whose
moduli space is expected to match that of an appropriate quiver
gauge theory, although we have not determined this gauge theory
in the present paper.

\subsec{Joining like cycles}

Let us put together $N$ D4-branes wrapped on degree $n$
surfaces. We propose that the resulting degree $nN$ surface
$P_{N|n}$ is given by \eqn\pmk{ \det \Phi^{(n)}(z^A)=0 } where
$\Phi^{(n)}$ is an $N\times N$ matrix of degree $n$ polynomials in
$z^A$. This geometry is described by the Higgs branch of an ${\cal N}=1$
$U(N)$ gauge theory with adjoint matter $\Phi_{(A_1\cdots A_n)}$.
The geometry of the D4-brane can be understood in terms of
a probe D0-brane, which introduces a superpotential of the form
\eqn\sdasup{ W = \phi_{04} \Phi_{(A_1\cdots A_n)}\phi_{40} z^{A_1}
\cdots z^{A_n} } Again, there are massless 0-4 open string modes when
the D0-brane coincides with the D4-brane, giving rise to the
condition \pmk.

As before, the divisor \pmk\ has a nice holomorphic curve $C$ given by
the vanishing of a row of minors in the matrix. Again we should
compute the intersection numbers $C\cdot C$ and $C\cdot J$. The
computation proceeds exactly as above, resulting in the integrals
\eqn\kcdj{ C\cdot J=\int_{{\bf P}^4 \times {\bf P}^{N-2}}
(1+x)(1+5x)(1+nx+y)^N=5 n^2 {N \choose 2},} and \eqn\kcdc{C\cdot
C=\int_{{\bf P}^4 \times {\bf P}^{N-3}} (1+5x)(1+nx+y)^N= 5 n^3 {N
\choose 3}.}

The D4-branes wrapped on the degree $n$ surfaces with a gauge field
$F=\kappa J$ (where $\kappa$ is integral or half integral according
to the parity of $n$) have the following charges \eqn\kdtc{q_2 = 5
\kappa n,~~~q_0 = - {5 \over 2} n \kappa^2 - {5\over 24} (n^3 + 10
n).}

The D4-brane wrapped on the degree $n N$ determinantal surface with
gauge field \eqn\kgf{F = C - {n(N-1) \over 2}J + \kappa J} has
charges \eqn\kdfc{q_2 = 5 n \kappa N,~~~ q_0 = - \Big[{5 \over 2} n
\kappa^2 + {5\over 24} (n^3 + 10 n)\Big] N.} in perfect agreement.

\subsec{Generic cycles}

Now we would like to analyze the case of $N$ D4-branes
wrapping cycles of degree $n_i$, $i=1, \ldots N$,  and carrying flux
$F=\kappa_i J$. We propose that a bound state of these D-branes is
again described by a cycle of the form

\eqn\gendet{\det \Phi(z) =0} where now the degree of the matrix
components is $\deg \Phi_{ij} = \half (n_i + n_j) + \kappa_i -
\kappa_j$. This 4-cycle has again a nontrivial, integral,
anti-self-dual form $F^-$, and the composite D4-brane carries a flux
$F = F^- + \kappa J$, where $\kappa= {\sum_i \kappa_i n_i \over
\sum_i n_i}$. If $\kappa_i \neq \kappa_j$ the component D4-branes
would generically preserve different supersymmetries, so the
existence of a supersymmetric bound state - that is, of a
supersymmetric vacuum for the $U(1)^N$ gauge theory - depends on the
K{\"a}hler moduli of the Calabi-Yau and is encoded in the D-term
constraints. A (piece of) convincing evidence for the correctness of
this construction is the nontrivial agreement in the D0-brane charge
as computed for the single D4-branes and for the bound state.
It would be interesting to understand the degrees of $\Phi_{ij}$ in terms of
$4-4'$ open string modes.

As before, the holomorphic curve $C_1$ is defined by setting the
first row of minors of $\Phi_{ij}$ to zero. Similar curves $C_i$,
defined by setting the $i^{th}$ row of minors to zero, differ from
$C_1$ by a multiple of $J$. The computation of the intersection
numbers $C_1 \cdot J $ and $C_1 \cdot C_2$ is slightly more subtle
than before.

Let us first calculate $C_1\cdot J$. We need to count solutions of
equations \eqn\asbef{\eqalign{& U(z)=0,~~~~H(z)=0, \cr & \sum_{2\leq
i \leq N} a_i \Phi_{ij}(z)  = 0,~~~~{1\leq j\leq N}  } } with the
identifications $a_i \sim \lambda a_i$ and $(z^A, a_i) \sim (\mu
z^A, \mu^{-\kappa_i-\half n_i} a_i )$. These identifications define a
bundle $W$ of ${\bf P}^{N-2}$ fibered over ${\bf P}^4$, which can
also be thought of as the projectivization of the vector bundle $V =
\bigoplus_{i=2}^{N} {\cal O}(-\kappa_i-\half n_i)$. The cohomology
ring has two generators, given by $x=\pi^* H_{{\bf P}^4}$ and
$y=-c_1(S)$ where $S$ is the universal subbundle of $\pi^*V$, and $\pi$ is
the projection map of the bundle. There are cohomology ring
relations $x^5=0$ and $y^{N-1} + \sum_k c_k(V) y^{N-1-k}=0$
\bott. Using \eqn\topdc{ \int_W x^4 y^{N-2}=1 } we can derive the
following relations \eqn\topints{ \eqalign{ & \int_W x^3
y^{N-1}=-c_1(V),\cr & \int_W x^2 y^{N}=-c_2(V)+c_1(V)^2,\cr & \int_W
x y^{N+1} = -c_3(V)+2c_1(V) c_2(V)-c_1(V)^3. } }

Now the problem of computing $C_1\cdot J$ amounts to counting the
zeros of a section of the vector bundle $L_x^5\oplus L_x\oplus
\bigoplus_{i=1}^N(L_x^{\half n_i-\kappa_i}\otimes L_y)$, where $L_x$
and $L_y$ are the line bundles over $W$ dual to $x$ and $y$
respectively. The answer is given by the Euler class \eqn\csdj{
\eqalign{ C_1\cdot J &= \int_{W} (1+5x)(1+x) \prod_{i=1}^N
\left(1+(-\kappa_i+\half n_i)x+y\right) \cr &=5\left[ \sum_{2\leq
i<j\leq N}n_in_j + \sum_{j=2}^N n_j \left(\kappa_j-\kappa_1 +
{n_1+n_j\over 2}\right) \right] } } where we have applied \topints.
Similarly we can compute
$C_1\cdot C_2$, using the vector bundle $V'=\bigoplus_{i=3}^N {\cal
O}(-\kappa_i-\half n_i)$ over ${\bf P}^4$. The problem again reduces
to computing the Euler class of $L_x^5 \oplus \bigoplus_{i=1}^N(L_x^{\half
n_i-\kappa_i}\otimes L_y)$, given by \eqn\eulsa{ \eqalign{ C_1\cdot
C_2 &= \int_{W'} (1+5x) \prod_{i=1}^N \left(1+(-\kappa_i+\half
n_i)x+y\right) \cr &=5 \left[ \sum_{3\leq i<j<k\leq N} n_in_jn_k +
\sum_{3\leq i<j\leq N} n_in_j\left( -\kappa_1-\kappa_2 +
{n_1+n_2\over 2} \right) \right. \cr & ~~+\sum_{3\leq i<j\leq
N}n_in_j\left( {n_i+n_j\over 2}+\kappa_i+\kappa_j \right) +
\sum_{j=3}^N n_j(\kappa_j+{n_j\over
2})\left(-\kappa_1-\kappa_2+{n_1+n_2\over 2}\right) \cr &~~\left.
+\sum_{j=3}^N n_j (-\kappa_1+{n_1\over 2})(-\kappa_2+{n_2\over 2}) +
\sum_{j=3}^N n_j(\kappa_j+{n_j\over 2})^2 \right] } } where $W'$ is
the projectivization of $V'$.

In fact, the homology classes of all the $C_i$'s are related to the same curve $C$
by \eqn\cresl{ C_i = C-(\kappa_i+{n_i\over 2})J,~~~~1\leq i\leq
N. } $C$ can be constructed as
\eqn\ccurv{ C:~~~ \sum_i v_i(z) \tilde \Phi_{ij}(z)=0,~~~~1\leq j\leq N, }
where $v_i(z)$ are homogeneous polynomials of degree $\half n_i+\kappa_i$.
In particular one can deform $C$ continuously so that only one of the $v_i$'s is nonzero,
and hence prove \cresl. Indeed one can check that the intersection numbers
\csdj\ and \eulsa\ are consistent with the relation \cresl. $C$ has intersection numbers
\eqn\intscsc{ \eqalign{ & C\cdot J = 5\left[ {1\over 2} \left(\sum
n_i\right)^2 + \sum n_i\kappa_i \right] \cr & C\cdot C =
5\left[{1\over 6}\left(\sum n_i\right)^3 + {1\over 12}\sum n_i^3 +
\sum n_i\sum n_j\kappa_j+\sum_i n_i\kappa_i^2 \right] }}

We shall consider the flux \eqn\ffsf{ F = C-{1\over 2}\left({\sum
n_i}\right)J. } With $F$ turned on, the D4-brane wrapped on the
surface defined by \gendet\ has induced D2 and D0-brane charges
\eqn\dtazffd{ \eqalign{ & q_2 = \sum n_i \kappa_i, \cr & q_0 = -
{5\over 24} \sum n_i^3 -{5\over 2}\sum n_i\kappa_i^2 -{25\over
12}\sum n_i } } which are precisely as expected from the $N$ D4-branes wrapped on
degree $n_i$ cycles.

\bigskip

\centerline{\bf Acknowledgements} We are grateful to S. Gukov, D. Jafferis,
G. Moore and C. Vafa for helpful discussions. This work is
supported in part by DOE grant DE-FG02-91ER40654.

\listrefs
\end